\documentclass[twocolumn,prd,superscriptaddress,showpacs]{revtex4-2}
\usepackage{graphicx}
\usepackage{dcolumn}
\usepackage{amsmath}

\begin{document}

\title{Scattering of low energy neutrinos and antineutrinos by neon and argon}
\author{Ian B. Whittingham}
\affiliation{College of Science and Engineering,
James Cook University, Townsville, Queensland, Australia 4811}

\date{\today}

\begin{abstract}
The theory of scattering of low energy neutrinos and antineutrinos by atomic 
electrons has recently been developed~(I. B. Whittingham, Phys. Rev. D $\mathbf{105}$ 013008 (2022)) 
using the Bound Interaction Picture in
configuration space to fully implement the relationship between the neutrino helicities and 
the orbital and spin angular momenta of the atomic electrons.
The energy spectra of ionization electrons produced by scattering of neutrinos and antineutrinos
with energies of 5, 10, 20, and 30 keV by hydrogen, helium and neon were calculated 
using Dirac screened Coulombic eigenfunctions.
This paper reports further applications of this theory, to a new calculation of the energy 
spectra for neon, as the original  calculation used some screening constants which 
underestimated the effects of screening in the inner subshells, and to scattering by argon.
The results are presented as ratios to the corresponding quantities for scattering by $Z$ free electrons.
The new spectra ratios for neon are larger than the original ratios by 
$\approx 0.03$ to $\approx 0.14$, with the greatest increases occurring for 10 keV neutrinos
and antineutrinos.
Integrated spectra ratios range from 0.16 to 0.59 for neon, and from 0.15 to 0.48 for argon, 
as the neutrino energy increases from 5 to 30 keV.
\end{abstract}

\maketitle

\section{Introduction}

There is significant interest in the low energy O(10 keV) scattering of electron neutrinos 
and antineutrinos by atomic electrons
\begin{equation}
\label{W0a}
\nu_{e} (\bar{\nu}_{e}) + e^{-} \rightarrow \nu_{e} (\bar{\nu}_{e}) + e^{-} ,
\end{equation}
for which the binding of the atomic electron cannot be ignored and one can expect modifications 
of the free electron scattering formulae. One example is the study of possible electromagnetic
properties of neutrinos, such as magnetic and electric dipole moments, using low energy 
scattering of neutrinos and antineutrinos~\cite{Giunti2015,Jeong2021}. For a review of
neutrino-atom collisions, see~\cite{Kouzakov2014}.

Of particular interest is the ionization of atoms by neutrinos and antineutrinos. Ionization 
cross sections calculated for H, He and Ne
were found~\cite{Gounaris2002} to be smaller than the corresponding free electron
cross sections, and the calculations were then extended~\cite{Gounaris2004} to the electron
spectra for H, He and Ne, and integrated ionization cross sections for H, He, Ne and Xe.

These calculations treat the $\nu_{e}$-electron scattering process as a probability weighted 
scattering by a free electron of mass $\tilde{m}$, where 
\begin{equation}
\label{W03a}
\tilde{m}^{2} = E_{e_{i}}^{2}-\mathbf{p}_{e_{i}}^{2},
\end{equation}
$E_{e_{i}} =m_{e}+\epsilon$ is the energy of the initial bound electron, where $\epsilon$ is the binding energy, 
and the momentum $\mathbf{p}_{e_{i}}$ is determined by the probability amplitude 
$|\Psi_{n_{i}l_{i}m_{i}}(\mathbf{p}_{e_{i}})|^{2}$, where
$\Psi_{n_{i}l_{i}m_{i}}(\mathbf{p}_{e_{i}})$ is the momentum-space atomic wave function.
Spin-independent non-relativistic atomic wave functions are used for the initial electron and Coulombic effects
on the final electron are ignored. These calculations destroy the relationship 
between the neutrino helicities and the orbital and spin angular momenta of the atomic electrons. 
Some of these issues have been addressed by~\cite{Chen2014,Chen2015,Chen2017} and their approach is
closest in spirit to the present calculations.

The theory of scattering of neutrinos and antineutrinos by bound electrons which maintains the full
collision dynamics has recently been developed~\cite{Whitt2022}. The scattering is 
formulated in configuation space using the Bound Interaction Picture~\cite{Furry1951,Whittingham1971}, 
rather than the usual formulation in the Interaction Picture in momentum space as appropriate to 
scattering by free electrons. 
The energy spectra of ionization electrons produced by scattering of neutrinos and antineutrinos
with energies of 5, 10, 20, and 30 keV by hydrogen, helium and neon were calculated 
using Dirac screened Coulombic eigenfunctions. The results were significantly different to those
of~\cite{Gounaris2004} and indicated that binding effects from both the initial bound state and
the final continuum state are important.

Very recently, a second quantization formalism has been developed~\cite{GPS2022} to include atomic
effects in the electron recoil signal for dark matter or neutrino scattering. Although the 
neutrino scattering amplitude includes more general interactions than the $W$- and $Z$- exchange
of the Standard Model, the formalism treats the atomic electrons as non-relativistic and includes
the electron spin through a non-relativistic approximation to the four-component Dirac 
eigenfunction which ignores coupling between the spin and orbital angular momenta.

This paper reports further applications of the Bound Interaction Picture theory~\cite{Whitt2022},
to a new calculation of the energy 
spectra for neon, as the original  calculation used some screening constants which 
underestimated the effects of screening in the inner subshells, and to scattering by argon as
it is the basis of current and near-future experiments sensitive to neutrino-electron scattering.

The general formalism for the scattering of neutrinos and antineutrinos by atomic electrons 
is summarized in Sec. II, and the radial matrix elements which occur in the atomic electron
scattering tensor discussed in Sec. III. Results for the energy spectra of the ionization
electrons produced in scattering from neon and argon are presented and discussed in Sec. IV,
and Sec. V contains a summary and conclusions for the investigation.

\section{General formalism for neutrino scattering by atomic electrons}

The scattering of neutrinos by atomic electrons involves both $W$- and $Z$- boson exchange. The
total $S$-matrix in the Bound Interaction Picture for $\nu_{e}$ scattering at low momentum transfers 
$k^{2} \ll M_{A}^{2}$, where $A=W,Z$, is~\cite{Whitt2022}   
\begin{equation}
\label{W34}
S_{fi}^{(\nu)}= -\pi i\frac{G_{\mathrm{F}}}{\sqrt{2}} \delta(E_{fi}^{(\nu)}) 
M^{(e)}_{n_{f},n_{i}}(\mathbf{q})^{\alpha} M^{(\nu)}(\mathbf{p}_{\nu_{f}},s_{f},\mathbf{p}_{\nu_{i}},s_{i})_{\alpha}
\end{equation}
where the atomic electron scattering amplitude is
\begin{eqnarray}
\label{W35}
M^{(e)}_{n_{f},n_{i}}(\mathbf{q})^{\alpha} & = & \int d^{3}x \;
e^{i(\mathbf{p}_{\nu_{i}}-\mathbf{p}_{\nu_{f}})\cdot \mathbf{x}}  
\nonumber  \\
&& \times
\bar{\phi}^{(+)}_{n_{f}} (\mathbf{x}) \gamma^{\alpha}(\bar{v}_{e}+\bar{a}_{e}\gamma_{5})
\phi^{(+)}_{n_{i}}(\mathbf{x}),
\end{eqnarray}
the quantity
\begin{equation}
\label{W29a}
\delta(E_{fi}^{(\nu)}) \equiv \delta(E_{n_{f}}+E_{\nu_{f}}-E_{n_{i}}-E_{\nu_{i}}) .
\end{equation}
incorporates energy conservation, $\mathbf{q}=\mathbf{p}_{\nu_{i}}-\mathbf{p}_{\nu_{f}}$ is 
the momentum transfer from the neutrino, and
\begin{equation}
\label{W36}
M^{(\nu)}(\mathbf{p}_{\nu_{f}},s_{f},\mathbf{p}_{\nu_{i}},s_{i})_{\alpha} = \bar{u}^{(s_{f})}(\mathbf{p}_{\nu_{f}})\gamma_{\alpha}
(1-\gamma_{5}) u^{(s_{i})}(\mathbf{p}_{\nu_{i}})
\end{equation}
is the neutrino scattering amplitude.
Here $\phi^{(+)}_{n}(\mathbf{x})$ is the positive energy eigenfunction for an electron in a 
state of the external field $A^{(\mathrm{ext})}$ specified by the quantum numbers $n$, and
$u^{(s)}(\mathbf{p}_{\nu})$ are the plane wave spinors describing a neutrino 
with momentum $\mathbf{p}_{\nu}$ and helicity $s$. 
The natural unit system $\hbar = c=1$ is used throughout, the 
scalar product of two 4-vectors is $A\cdot B \equiv g^{\alpha \beta} A_{\alpha} B_{\beta} 
= A_{0}B_{0}-\mathbf{A}\cdot \mathbf{B}$, the Dirac matrices 
$\gamma^{\alpha}, (\alpha =0,1,2,3)$ satisfy $\{\gamma^{\alpha},\gamma^{\beta}\}=2g^{\alpha \beta}$,
and $\gamma_{5}\equiv i\gamma^{0}\gamma^{1}\gamma^{2}\gamma^{3}$.

The electron mixing parameters are
\begin{equation}
\label{W37}
\bar{v}_{e} = v_{e}+2 , \quad \bar{a}_{e}= a_{e}-2,
\end{equation}
where\begin{equation}
\label{W37a}
v_{e}= -1+4 \sin^{2} \theta_{W}, \quad a_{e}= 1,
\end{equation}
and $\theta_{W}$ is the weak mixing angle.

For scattering of antineutrinos, $M^{(\nu)}$ is replaced by
\begin{equation}
\label{W36a}
M^{(\bar{\nu})}(\mathbf{p}_{\nu_{f}},s_{f},\mathbf{p}_{\nu_{i}},s_{i})_{\alpha} = 
\bar{v}^{(s_{i})}(\mathbf{p}_{\nu_{i}})\gamma_{\alpha}
(1-\gamma_{5}) v^{(s_{f})}(\mathbf{p}_{\nu_{f}}),
\end{equation}
where $v^{(s)}(\mathbf{p}_{\nu})$ is the antineutrino plane wave spinor,
and $\delta(E_{fi}^{(\nu)})$ is replaced by $\delta(E_{fi}^{(\bar{\nu})})$.

We assume the atomic electron moves in a spherically symmetric potential 
$V(r)=e A^{(\mathrm{ext})}(r)$, for which the Dirac equation has eigenfunctions 
of the form~\cite{Rose1961}
\begin{equation}
\label{W45}
\phi_{\kappa, \mu,E}(r, \theta, \varphi ) = \frac{1}{r} \left( 
\begin{array}{l}
g_{\kappa,E}(r) \chi^{\mu}_{\kappa}(\Omega) \\
i f_{\kappa, E} (r) \chi^{\mu}_{-\kappa}(\Omega)
\end{array}
\right).
\end{equation}
where $(r, \theta, \varphi )=(r, \Omega)$ are spherical polar coordinates, and $\chi^{\mu}_{\kappa}(\Omega)$
are the spinor spherical harmonics
\begin{equation}
\label{W46}
\chi^{\mu}_{\kappa}(\Omega) = \sum_{m_{s}} C(l_{\kappa},\frac{1}{2},j,\mu-m_{s},m_{s},\mu ) 
Y^{\mu-m_{s}}_{l_{\kappa}}(\Omega) \chi_{m_{s}}.
\end{equation}
Here $C(j_{1},j_{2},j_{3},m_{1},m_{2},m_{3})$ is a Clebsch-Gordon coefficient, and
$\chi_{m_{s}}$ are the two component Pauli spinors. The total angular momentum $j$ and orbital angular 
momentum $l_{\kappa}$ are obtained from the quantum number $\kappa$ by
\begin{equation}
\label{W47}
j=|\kappa|-\frac{1}{2}, l_{\kappa}=\left\{ \begin{array}{cl}
\kappa & \kappa > 0  \\
-\kappa-1& \kappa < 0
\end{array}, \right.
l_{-\kappa} = l_{\kappa} - \frac{\kappa}{|\kappa|},
\end{equation}
where $\kappa$ takes all non-zero integral values. 
The radial functions satisfy
\begin{equation}
\label{W48}
\left( \begin{array}{cc}
d/dr + \kappa /r & -(E+m_{e}-V(r))  \\
E-m_{e}-V(r) & d/dr-\kappa /r
\end{array}
\right)
\left( \begin{array}{c}
g_{\kappa,E}(r) \\
f_{\kappa,E}(r) 
\end{array}
\right)  = 0.
\end{equation}

The energy spectrum of the ionized electrons is~\cite{Whitt2022}
\begin{equation}
\label{W70a}
\frac{d \sigma^{(\nu)} }{d E_{f}} =  \frac{G_{\mathrm{F}}^{2}}{8\pi}
\frac{1}{16 m_{e} E_{\nu_{i}}^{2}}  \int  d q^{2} 
L_{fi}(\tilde{\mathbf{q}},p_{\nu_{i}},p_{\nu_{f}}),
\end{equation}
where
\begin{eqnarray}
\label{W69b}
L_{fi}(\tilde{\mathbf{q}},p_{\nu_{i}},p_{\nu_{f}}) & = &
\mathrm{Re}[L^{(e)}_{fi}(\tilde{\mathbf{q}})^{\beta \alpha}]
\mathrm{Re}[ L^{(\nu)}(p_{\nu_{i}},p_{\nu_{f}})_{\beta \alpha} ]  \nonumber  \\
&& -
\mathrm{Im}[L^{(e)}_{fi}(\tilde{\mathbf{q}})^{\beta \alpha}]
\mathrm{Im}[L^{(\nu)}(p_{\nu_{i}},p_{\nu_{f}})_{\beta \alpha}] \nonumber  \\
\end{eqnarray}
and it is understood that $E_{\nu_{f}}=E_{i}+E_{\nu_{i}}-E_{f}$.
The neutrino scattering tensor is 
\begin{eqnarray}
\label{W44}
L^{(\nu)}(p_{\nu_{i}},p_{\nu_{f}})^{\beta \alpha} &\equiv & 
[\bar{u}^{(s_{f})}(\mathbf{p}_{\nu_{f}})\gamma^{\beta}(1-\gamma_{5}) u^{(s_{i})}(\mathbf{p}_{\nu_{i}})]^{\dagger}
\nonumber  \\
&& \times \bar{u}^{(s_{f})}(\mathbf{p}_{\nu_{f}})\gamma^{\alpha}(1-\gamma_{5}) u^{(s_{i})}(\mathbf{p}_{\nu_{i}})
\nonumber \\
& = & 8 (p_{\nu_{i}}^{\beta}\,p_{\nu_{f}}^{\alpha} + p_{\nu_{i}}^{\alpha}\, p_{\nu_{f}}^{\beta} 
- p_{\nu_{i}} \cdot p_{\nu_{f}}\, g^{\beta \alpha}
\nonumber  \\
&& + i \epsilon^{\rho \beta \lambda \alpha}\, p_{\nu_{i,\rho}}\, p_{\nu_{f, \lambda}})  \label{W44a},
\end{eqnarray}
where $s_{i}=s_{f}=-1/2$.
The atomic electron scattering tensor is 
\begin{eqnarray}
\label{W68}
L^{(e)}_{fi}(\tilde{\mathbf{q}})^{\beta \alpha} & = & \sum_{\kappa_{f},\bar{l},l} i^{l-\bar{l}} 
(2\bar{l}+1)(2l+1) \left[
\bar{v}_{e}^{2} L^{\beta \alpha}_{v_{e}v_{e}} \right.
\nonumber  \\
&& \left. +\bar{a}_{e}^{2} L^{\beta \alpha }_{a_{e}a_{e}}
+ \bar{v}_{e}\bar{a}_{e} (L^{\beta \alpha}_{v_{e}a_{e}} +L^{\beta \alpha}_{a_{e}v_{e}})\right],
\end{eqnarray} 
where $\tilde{\mathbf{q}} \equiv (0,0,q)$ and $q \equiv |\mathbf{q}|$.
The quantities $L^{\beta \alpha}_{v_{e}v_{e}}$, etc 
involve~\cite{Whitt2022} various angular momentum coupling coefficients and the radial integrals 
\begin{eqnarray}
\label{W53}
I^{gg}_{l}(q) &\equiv & \int dr \,g^{*}_{\kappa_{f},E_{f}}(r) j_{l}(qr) g_{\kappa_{i},E_{i}}(r),
\nonumber \\
I^{gf}_{l}(q) &\equiv & \int dr \,g^{*}_{\kappa_{f},E_{f}}(r) j_{l}(qr) f_{\kappa_{i},E_{i}}(r),
\nonumber  \\
I^{fg}_{l}(q) &\equiv & \int dr \,f^{*}_{\kappa_{f},E_{f}}(r) j_{l}(qr) g_{\kappa_{i},E_{i}}(r),
\nonumber \\
I^{ff}_{l}(q) &\equiv & \int dr \,f^{*}_{\kappa_{f},E_{f}}(r) j_{l}(qr) f_{\kappa_{i},E_{i}}(r).
\end{eqnarray}

The scattering of antineutrinos involves
\begin{eqnarray}
\label{W44a}
L^{(\bar{\nu})}(p_{\nu_{i}},p_{\nu_{f}})^{\beta \alpha} &\equiv & 
[\bar{v}^{(s_{i})}(\mathbf{p}_{\nu_{i}})\gamma^{\beta}(1-\gamma_{5}) v^{(s_{f})}(\mathbf{p}_{\nu_{f}})]^{\dagger}
\nonumber  \\
&& \times \bar{v}^{(s_{i})}(\mathbf{p}_{\nu_{i}})\gamma^{\alpha}(1-\gamma_{5}) v^{(s_{f})}(\mathbf{p}_{\nu_{f}})
\nonumber \\
& = & L^{(\nu)}(-p_{\nu_{f}},-p_{\nu_{i}})^{\beta \alpha},  
\nonumber \\
& = & [L^{(\nu)}(p_{\nu_{i}},p_{\nu_{f}})^{\beta \alpha}]^{*},
\end{eqnarray}
where $v^{(s)}(\mathbf{p}_{\nu})$ is the antineutrino plane wave spinor,
and $s_{i}=s_{f}=+1/2$.

\section{Radial matrix elements}

The radial integrals (\ref{W53}) involve the Dirac radial functions $g_{\kappa,E}(r)$ 
and $f_{\kappa,E}(r)$ for the initial bound electron and the final continuum electron. 
We assume a potential of a Coulombic form $V(r)=-\alpha Z_{\mathrm{eff}} /r$, where 
$Z_{\mathrm{eff}}$ is an effective nuclear charge. 
The radial Dirac equations (\ref{W48}) then have analytic 
solutions~\cite{Rose1961} in terms of confluent hypergeometric functions ${}_{1}F_{1}(a,c,z)$.

As scattering by electrons in the ground states of Ne and Ar involve only K-, L- and M- shell 
electrons, we can use the simplified expressions~\cite{Rose1961,Das1973}
\begin{eqnarray}
\label{W71}
\left ( \begin{array}{cc}
g_{\kappa_{i},E_{i}}(r) \\
f_{\kappa_{i},E_{i}}(r) 
\end{array}
\right)
& = &
N_{i} \left( \begin{array}{cc}
\sqrt{m_{e}+E_{i}} \\
- \sqrt{m_{e}-E_{i}} 
\end{array}
\right) (2 \lambda_{i} r)^{\gamma_{i}}e^{-\lambda_{i}r} 
\nonumber \\
&& \times \left[ \left( \begin{array} {cc}
c_{0} \\
a_{0} 
\end{array} \right) + \left( \begin{array} {cc}
c_{1} \\
a_{1}
\end{array}  \right) \lambda_{i} r \right.
\nonumber  \\
&& \left. + \left( \begin{array} {cc}
c_{2} \\
a_{2}
\end{array}  \right) (\lambda_{i} r)^{2} \right]
\end{eqnarray}
where
\begin{equation}
\label{W72}
\lambda_{i} \equiv \sqrt{m_{e}^{2}-E_{i}^{2}}, \quad \gamma_{i} \equiv \sqrt{\kappa_{i}^{2}-\zeta^{2}},
\quad \zeta \equiv \alpha Z_{\mathrm{eff}}   ,
\end{equation}
and the initial state energy $E_{i}$ is
\begin{equation}
\label{W73}
E_{n,\kappa_{i}}= m_{e}\left[1 +\left(\frac{\zeta}{n-|\kappa_{i}|+\gamma_{i}} \right)^{2} \right]^{-1/2}.
\end{equation}
The dimensionless coefficients $(c_{i},a_{i}, N_{i})$ for the K-shell ($n=1, \kappa_{i}=-1$), 
L$_{\mathrm{I}}$ - subshell ($n=2, \kappa_{i}=-1$), L$_{\mathrm{II}}$ -subshell ($n=2, \kappa_{i}=+1$),
L$_{\mathrm{III}}$ -subshell ($n=2,\kappa_{i}=-2$), 
M$_{\mathrm{I}}$ - subshell ($n=3, \kappa_{i}=-1$), M$_{\mathrm{II}}$ -subshell ($n=3, \kappa_{i}=+1$),
and M$_{\mathrm{III}}$ -subshell ($n=3,\kappa_{i}=-2$) are listed in Table I.

\begin{table*}
\caption{\label{tab:bound}Parameters defining the K-, L- and M-shell Coulombic radial eigenfunctions for Ne and Ar, 
expressed in terms of $\zeta = \alpha Z$ and $\gamma =\sqrt{\kappa^{2}-\zeta^{2}}$.}
\begin{ruledtabular}
\begin{tabular}{cccccccccc}
Subshell & $E_{n \kappa} $ &  $\lambda$ & $N_{i}$ & $a_{0}$ & $c_{0}$ & $a_{1}$ & $c_{1}$ & $a_{2}$ & $c_{2}$ \\
\hline
$K$ & $\gamma $ & $\zeta $ & $\left( \frac{\zeta }{\Gamma (2 \gamma +1)}\right)^{1/2}$ & 
$1$ & $1$ & $0$ & $0$ & $0$ & $0$  \\
$L_{\mathrm{I}}$ & $ \left(\frac{\gamma +1}{2}\right)^{1/2}$ & $\frac{\zeta}{2E} $ & 
$\frac{1}{2}\left[\frac{\lambda (2 \gamma +1)}{E (2E+1)\Gamma (2 \gamma +1)}\right]^{1/2} $ &
$2(E+1)$ & $2E$ & $-2\left(\frac{2E+1}{2 \gamma +1}\right) $ & $-2\left(\frac{2E+1}{2 \gamma +1}\right) $ & $0$ & $0$ \\
$L_{\mathrm{II}}$ & $ \left(\frac{\gamma +1}{2}\right)^{1/2}$ & $\frac{\zeta}{2E} $ & 
$\frac{1}{2}\left[\frac{\lambda (2 \gamma +1)}{E (2E-1)\Gamma (2 \gamma +1)}\right]^{1/2} $ &
$2E$ & $2(E-1)$ & $-2\left(\frac{2E-1}{2 \gamma +1}\right) $ & $-2\left(\frac{2E-1}{2 \gamma +1}\right) $ & $0$ & $0$ \\
$L_{\mathrm{III}}$ & $ \frac{\gamma}{2}$ & $ \frac{\zeta}{2}$ & $\left(\frac{\lambda}{\Gamma (2 \gamma +1)}\right)^{1/2} $ & 
$1$ & $1$ & $0$ & $0$ & $0$ & $0$  \\
$M_{\mathrm{I}}$ & $\frac{\gamma +2}{\sqrt{4 \gamma +5}}$ & $\frac{\zeta}{\sqrt{4 \gamma +5}}$ & 
$ \lambda \left[\frac{\lambda (\gamma +1) (2 \gamma +1)}{2 \zeta (\zeta + \lambda) \Gamma (2 \gamma +1)}\right]^{1/2} $ &
$ 3+ \frac{\zeta}{\lambda}$ & $ -1+\frac{\zeta}{\lambda}$ & $ -\frac{4}{2 \gamma +1} \left(2 + \frac{\zeta}{\lambda }\right) $ &
$ -\frac{4 \zeta}{\lambda (2 \gamma +1)} $ & $ \frac{2}{(\gamma+1)(2 \gamma +1)}\left(1+\frac{\zeta}{\lambda}\right) $
& $ a_{2}$ \\
$M_{\mathrm{II}}$ & $\frac{\gamma +2}{\sqrt{4 \gamma +5}}$ & $\frac{\zeta}{\sqrt{4 \gamma +5}}$ & 
$ \lambda \left[\frac{\lambda (\gamma +1) (2 \gamma +1)}{2 \zeta (\zeta - \lambda) \Gamma (2 \gamma +1)}\right]^{1/2} $ &
$1+\frac{\zeta}{\lambda} $ & $ -3+\frac{\zeta}{\lambda}$ & 
$ -\frac{4 \zeta}{\lambda (2 \gamma +1)} $ &
$ \frac{4}{2 \gamma +1} \left(2 - \frac{\zeta}{\lambda }\right) $ &
$ -\frac{2}{(\gamma+1)(2 \gamma +1)}\left(1-\frac{\zeta}{\lambda}\right) $
& $ a_{2}$ \\
$M_{\mathrm{III}}$ & $\frac{\gamma +1}{\sqrt{2 \gamma +5}}$ & $\frac{\zeta}{\sqrt{2 \gamma +5}}$ & 
$ \lambda \left[\frac{\lambda (2 \gamma +1)}{2 \zeta (\zeta + 2\lambda) \Gamma (2 \gamma +1)}\right]^{1/2} $ &
$3+\frac{\zeta}{\lambda} $ & $ 1+\frac{\zeta}{\lambda}$ & 
$ -\frac{2}{2 \gamma +1} \left(2 + \frac{\zeta}{\lambda }\right) $ &
$ -\frac{2}{2 \gamma +1} \left(2 + \frac{\zeta}{\lambda }\right) $ & $0$ & $0$ \\
\end{tabular}
\end{ruledtabular}
\end{table*}

The screening effects of the electrons in the filled K- shell, L-subshells and M-subshells are 
represented by an effective nuclear charge of the form $Z_{\mathrm{eff}}=Z-s_{i}$,
a procedure that should be reasonable for small principal quantum number $n$ and small 
$n-l$~\cite{Bethe1957}. These screening constants $s_{i}$ are taken from the 
fits~\cite{Thomas1997} of Dirac single electron eigenfunctions to empirical binding energies.
For neon, they are
2.016 (K-shell), 6.254 (L$_{\mathrm{I}}$ -subshell), and 7.482 (L$_{\mathrm{II}}$- and 
L$_{\mathrm{III}}$- subshells).
For argon, the constants are
2.721 (K-shell), 8.248 (L$_{\mathrm{I}}$ -subshell), 
9.470 (L$_{\mathrm{II}}$- and L$_{\mathrm{III}}$- subshells),
13.603 (M$_{\mathrm{I}}$ -subshell), and 14.771 (M$_{\mathrm{II}}$- and M$_{\mathrm{III}}$- subshells).

The final electron continuum states involve the hypergeometric function ${}_{1}F_{1}(a,c,z)$
where $a = \gamma_{f}+1+iy$, $c=2 \gamma_{f}+1$, $z= 2ip_{f}r$,  and 
\begin{equation}
\label{W75}
\gamma_{f}\equiv \sqrt{\kappa_{f}^{2}-\zeta^{2}}, \quad p_{f} \equiv \sqrt{E_{f}^{2}-m_{e}^{2}},
\quad y \equiv \frac{\zeta E_{f}}{p_{f}}.
\end{equation} 
We choose to integrate the Dirac equation directly to obtain the continuum states 
rather than evaluate the hypergeometric functions as $a$ and $z$ are complex and
the computation of ${}_{1}F_{1}(a,c,z)$  would involve the 
summation of a slowly convergent complex series for each required value of $z$. 
For each shell and subshell calculation, the continuum state electrons are assumed to move in the 
same potential as the bound state electrons~\cite{Schofield1973}.

\section{Results and discussion}

The energy spectra $d \sigma/dE_{f}$ of the ionization electrons have been calculated as a 
function of the electron kinetic energy $\epsilon_{f} = E_{f}-m_{e}$.
Results are obtained for scattering of 5, 10, 20, and 30 keV neutrino energies
by the ground state systems Ne($1s^{2}2s^{2}2p^{6}$) and Ar($1s^{2}2s^{2}2p^{6}3s^{2}3p^{6}$), 
where the electrons are considered as independent scattering centers.

The energy spectra are compared to those for scattering from free electrons, 
for which~\cite{Gounaris2002}
\begin{eqnarray}
\label{W102}
\left(\frac{d \sigma^{(\nu)}}{d E_{f}}\right)_{(\mathrm{Free})}  & = & 
\frac{ G_{F}^{2} m_{e}}{8 \pi E_{\nu_{i}}^{2}} \left\{
(\bar{v}_{e}-\bar{a}_{e})^{2} E_{\nu_{i}}^{2} \right.
\nonumber  \\
&& + (\bar{v}_{e}+\bar{a}_{e})^{2}(E_{\nu_{i}}+m_{e}-E_{f})^{2}
\nonumber  \\
&& \left. + m_{e}(\bar{v}_{e}^{2}-\bar{a}_{e}^{2}) (m_{e}-E_{f}) \right\},
\end{eqnarray}
where $m_{e} \le E_{f} \le m_{e}+ \epsilon_{f}^{\mathrm{max}}$ and the maximum kinetic energy is
\begin{equation}
\label{W103}
\epsilon_{f}^{\mathrm{max}} = \frac{2 E_{\nu_{i}}^{2}}{m_{e}+2E_{\nu_{i}}}.
\end{equation}

The total cross section for scattering off free electrons is~\cite{Gounaris2002}
\begin{eqnarray}
\label{W102a}
\sigma^{(\nu)}_{(\mathrm{Free})}  & = & 
\frac{ G_{F}^{2} m_{e}E_{\nu_{i}}}{8 \pi } \left[
(\bar{v}_{e}-\bar{a}_{e})^{2} \frac{2 E_{\nu_{i}}}{m_{e}+2 E_{\nu_{i}}} \right.
\nonumber  \\
&& + \frac{1}{3}(\bar{v}_{e}+\bar{a}_{e})^{2}\left\{1-\frac{m_{e}^{3}}{(m_{e}+2 E_{\nu_{i}})^{3}}\right\}
\nonumber  \\
&& \left. - (\bar{v}_{e}^{2}-\bar{a}_{e}^{2}) \frac{2 m_{e} E_{\nu_{i}}}{(m_{e}+2 E_{\nu_{i}})^{2} }\right].
\end{eqnarray}

The expressions for $\bar{\nu}_{e}$ scattering can be obtained by the interchange
$\bar{a}_{e} \leftrightarrow -\bar{a}_{e}$ in (\ref{W102}) and
(\ref{W102a}).

Results for $\nu_{e}$ and $\bar{\nu}_{e}$ scattering by Ne are given in Table II, 
and by Ar in Table III. 
The energy spectra and cross sections are expressed as ratios 
\begin{equation}
\label{W100a}
R^{(\nu)}(E_{f}) = \frac{d\sigma ^{(\nu)}/dE_{f}}{Z(d\sigma ^{(\nu)}/dE_{f})_{(\mathrm{Free})}},
\end{equation}
and $\sigma^{(\nu)}/Z\sigma^{(\nu)}_{(\mathrm{Free})}$,
to the corresponding quantities for scattering by $Z$ free electrons. Tables and plots of the 
energy spectra for scattering from free electrons are given in~\cite{Whitt2022}.

\begin{table*}
\caption{\label{tab:nu-Ne}Energy spectra $d\sigma^{(\nu)} /dE_{f}$  of the ionization electrons from 
scattering of incident neutrinos of energy $E_{\nu_{i}}$ by neon. The results are 
expressed as ratios to the spectra $10(d\sigma^{(\nu)}/dE_{f}) _{(\mathrm{Free})}$  
for scattering by 10 free electrons and are given as a function of the kinetic energy $\epsilon_{f}$ of the electron . 
Also shown are the integrated spectra $\sigma^{(\nu)}$ expressed as a ratio to the integrated spectra
$10\sigma^{(\nu)}_{(\mathrm{Free})}$ for 10 free electrons.
Results for scattering of antineutrinos are given in parentheses. }
\begin{ruledtabular}
\begin{tabular}{lllll}
$\epsilon_{f}/\epsilon_{f}^{\mathrm{max}}$ & $E_{\nu_{i}}=5 $ (keV)&  $E_{\nu_{i}}=10$ (keV) &$ E_{\nu_{i}}=20$ (keV) & $E_{\nu_{i}}=30$ (keV) \\
\hline
0.01 & 0.2175 (0.2134) & 0.3867 (0.3821) & 0.4977 (0.4882) & 0.5720 (0.5578) \\
0.1  & 0.2041 (0.2004) & 0.3943 (0.3896) & 0.5258 (0.5171) & 0.6155 (0.6020) \\
0.2  & 0.1880 (0.1848) & 0.3983 (0.3933) & 0.5358 (0.5282) & 0.6272 (0.6148) \\
0.3  & 0.1716 (0.1690) & 0.3982 (0.3929) & 0.5383 (0.5320) & 0.6287 (0.6187) \\
0.4  & 0.1566 (0.1546) & 0.3930 (0.3876) & 0.5370 (0.5321) & 0.6249 (0.6168)   \\
0.5  & 0.1444 (0.1430) & 0.3794 (0.3745) & 0.5330 (0.5296) & 0.6170 (0.6121)   \\
0.6  & 0.1359 (0.1350) & 0.3524 (0.3490) & 0.5264 (0.5245) & 0.6055 (0.6044)   \\
0.7  & 0.1305 (0.1302) & 0.3102 (0.3093) & 0.5140 (0.5139) & 0.5901 (0.5934)   \\
0.8  & 0.1263 (0.1268) & 0.2636 (0.2655) & 0.4750 (0.4788) & 0.5666 (0.5748)   \\
0.9  & 0.1210 (0.1226) & 0.2300 (0.2347) & 0.3710 (0.3815) & 0.4792 (0.4957)   \\
1.0  & 0.1136 (0.1166) & 0.2021 (0.2108) & 0.2745 (0.2922) & 0.3051 (0.3302)   \\
$\epsilon_{f}^{\mathrm{max}}$ (eV) & 95.969 & 376.65 & 1451.9 & 3152.4  \\
$\sigma^{(\nu)}/\sigma^{(\nu)}_{(\mathrm{Free})}$ &  0.1612 (0.1600) &  0.3549 (0.3540) & 0.5051 (0.5050) & 0.5910 (0.5909) \\
\end{tabular}
\end{ruledtabular}
\end{table*}
\begin{table*}
\caption{\label{tab:nu-Ar}Energy spectra $d\sigma^{(\nu)} /dE_{f}$  of the ionization electrons from 
scattering of incident neutrinos of energy $E_{\nu_{i}}$ by argon. The results are 
expressed as ratios to the spectra $18(d\sigma^{(\nu)}/dE_{f}) _{(\mathrm{Free})}$  
for scattering by 18 free electrons and are given as a function of the kinetic energy $\epsilon_{f}$ of the electron . 
Also shown are the integrated spectra $\sigma^{(\nu)}$ expressed as a ratio to the integrated spectra
$18\sigma^{(\nu)}_{(\mathrm{Free})}$ for 18 free electrons.
Results for scattering of antineutrinos are given in parentheses.}
\begin{ruledtabular}
\begin{tabular}{lllll}
$\epsilon_{f}/\epsilon_{f}^{\mathrm{max}}$ & $E_{\nu_{i}}=5 $ (keV)&  $E_{\nu_{i}}=10$ (keV) &$ E_{\nu_{i}}=20$ (keV)& $E_{\nu_{i}}=30$ (keV) \\
\hline
0.01 & 0.1867 (0.1845) & 0.2855 (0.2813) & 0.4318 (0.4228) & 0.4964 (0.4859) \\
0.1  & 0.1875 (0.1851) & 0.2929 (0.2888) & 0.4493 (0.4404) & 0.5223 (0.5124)  \\
0.2  & 0.1856 (0.1830) & 0.2945 (0.2907) & 0.4502 (0.4417) & 0.5275 (0.5183)  \\
0.3  & 0.1792 (0.1767) & 0.2929 (0.2894) & 0.4451 (0.4374) & 0.5256 (0.5171)  \\
0.4  & 0.1669 (0.1647) & 0.2890 (0.2861) & 0.4361 (0.4297) & 0.5169 (0.5095)  \\
0.5  & 0.1493 (0.1478) & 0.2842 (0.2818) & 0.4240 (0.4192) & 0.5007 (0.4948)  \\
0.6  & 0.1311 (0.1301) & 0.2801 (0.2781) & 0.4096 (0.4069) & 0.4756 (0.4726)  \\
0.7  & 0.1170 (0.1166) & 0.2712 (0.2699) & 0.3917 (0.3918) & 0.4419 (0.4429)  \\
0.8  & 0.1067 (0.1068) & 0.2352 (0.2360) & 0.3693 (0.3730) & 0.4002 (0.4065)  \\
0.9  & 0.09465 (0.09562) & 0.1865 (0.1898) & 0.3251 (0.3339) & 0.3494 (0.3630)  \\
1.0  & 0.07822 (0.08011) & 0.1499 (0.1559) & 0.2164 (0.2316) & 0.2220 (0.2449)  \\
$\epsilon_{f}^{\mathrm{max}}$ (eV) & 95.969 & 376.65 & 1451.9 & 3152.4  \\
$\sigma^{(\bar{\nu})}/\sigma^{(\bar{\nu})}_{(\mathrm{Free})}$ &  0.1533 (0.1524) & 0.2730 (0.2719) & 0.4138 (0.4128) 
 & 0.4779 (0.4799)  \\
\end{tabular}
\end{ruledtabular}
\end{table*}

\begin{figure}[ht]
\includegraphics[width=0.95\columnwidth]{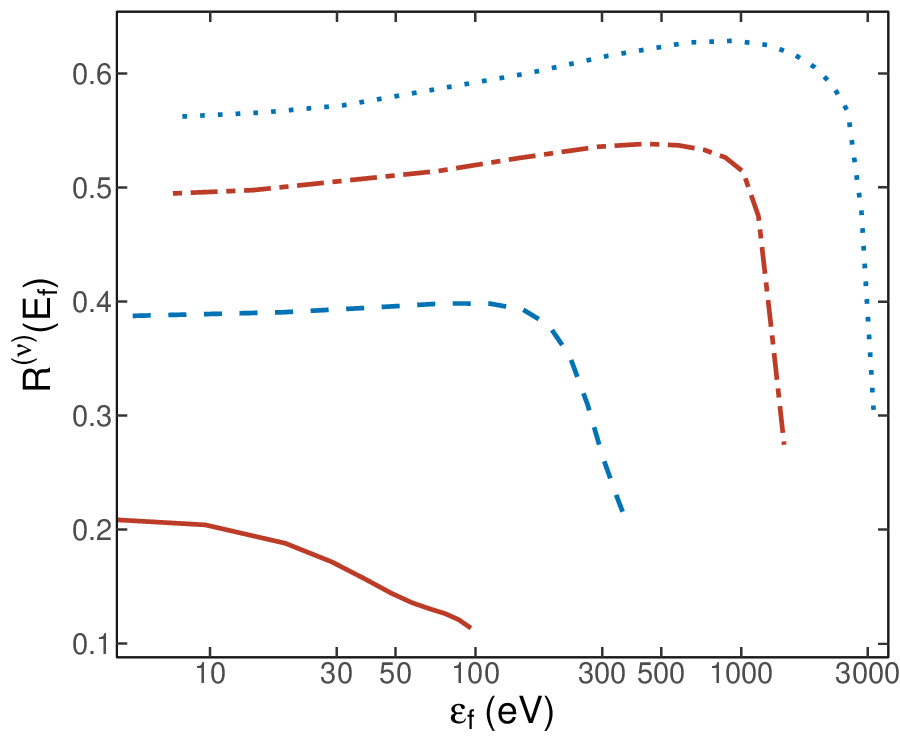}
\caption{\label{fig:Neon}Energy spectra ratios $R^{(\nu)}(E_{f})$ (Eqn (\ref{W100a})),  
as a function of electron kinetic energy $\epsilon_{f}$,
of ionization electrons resulting from scattering of neutrinos by ground state neon.
Results are shown for scattering of 5 keV (solid line), 10 keV (dashed line), 20 keV (dash-dotted line), 
and 30 keV (dotted line) incident neutrino energies.}
\end{figure}

\begin{figure}[ht]
\includegraphics[width=0.95\columnwidth]{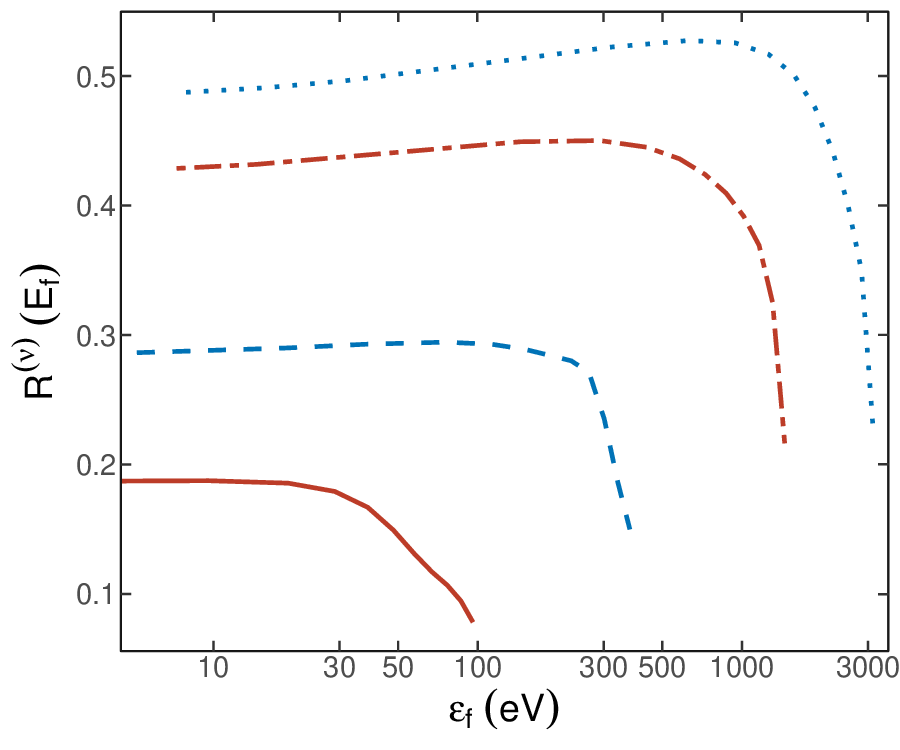}
\caption{\label{fig:Argon}Energy spectra ratios $R^{(\nu)}(E_{f})$ (Eqn (\ref{W100a})),  
as a function of electron kinetic energy $\epsilon_{f}$,
of ionization electrons resulting from scattering of neutrinos by ground state argon.
Results are shown for scattering of 5 keV (solid line), 10 keV (dashed line), 20 keV (dash-dotted line), 
and 30 keV (dotted line) incident neutrino energies.}
\end{figure}

The energy spectra ratios for scattering of neutrinos by Ne and Ar are shown in 
Figs \ref{fig:Neon} and \ref{fig:Argon} respectively. 
Plots for scattering of antineutrinos are not shown as they differ only very slightly 
from those for scattering by neutrinos. The energy spectra ratios become constant 
at low kinetic energies and can safely be extrapolated to lower kinetic energies.

The present spectra for Ne have the same general form as those calculated in \cite{Gounaris2004},
but have a much smoother structure and are significantly lower, representing a 
stronger dependence on atomic binding effects.

The largest binding energy of 3.2059 keV, for the K-shell of argon, is less than the lowest 
incident neutrino energy of $E_{\nu_{i}}$= 5 keV considered in the present calculations.
Consequently, all occupied shells and subshells contribute to the energy spectra although,
for the lower neutrino energies, the contributions from the inner shell and subshells will be
small.
For neon, the K-shell contributions at 5 keV and 10 keV are only O($10^{-3}$) and
O($10^{-2}$) respectively. At all neutrino energies, the L$_\mathrm{I}$ subshell gives the 
largest contribution, and the L$_\mathrm{II}$ and L$_\mathrm{III}$ contributions are quite similar. 
For argon, the K-shell contributions at 5 keV and 10 keV are O($10^{-5}$) and
O($10^{-3}$) respectively, and the L-subshell contributions at 5 keV are O($10^{-2}$).
The M$_\mathrm{I}$ subshell contribution is dominant at all energies, the L$_\mathrm{II}$ and 
L$_\mathrm{III}$ contributions are similar, as are the M$_\mathrm{II}$ and M$_\mathrm{III}$ contributions.

The major computational challenge is to obtain satisfactory convergence in the
sum over $\kappa_{f}$, as the convergence decreases with increasing $\epsilon_{f}$.  
The imposed practical limit $|\kappa_{f}| \le 50$ ensured the fractional contributions 
of the neglected terms were less 
than $1 \times 10^{-4}$ for $\epsilon_{f} \alt 1.4 $ keV in the Ne spectra, and for 
$\epsilon_{f} \alt 0.6 $ keV in the Ne spectra.
However, for the higher energies, the convergence decreased and the contributions 
of the neglected terms increased to $1 \times 10^{-3}$ 
for the L subshells of Ne, and $1 \times 10^{-2}$ for the M subshells of Ar, so the numbers shown for these  
higher energies are slight underestimates. 

The binding effects in the spectra increase with atomic number, 
decrease with increasing $E_{\nu_{i}}$ and, for each $E_{\nu_{i}}$, are most significant at the 
high electron energy end of the spectrum. The sharp decrease in the spectra at the high 
energy end is due to the minimizaion of the range $2E_{\nu_{f}}$  of the $q^2$ integration 
in (\ref{W70a}) at $E_{f}^{\mathrm{max}}$. 
Also, for this region, $q \approx E_{\nu_{i}}$, so that the 
high energy tail will increase as $E_{\nu_{i}}$ increases.

The integrated cross sections
\begin{equation}
\label{W109a}
\sigma^{(\nu)} = \int_{E_{l}}^{E_{u}} dE_{f}  \, \frac{d\sigma^{(\nu)} }{dE_{f}},
\end{equation}
where $E_{l}=m_{e}$ and $E_{u}= m_{e}+\epsilon_{f}^{\mathrm{max}}$, can be estimated from the 
calculated energy spectra. We assume the spectra at $E_{f}=m_{e}$
are the same as at $m_{e}+ 0.01 \epsilon_{f}^{\mathrm{max}}$. 
These integrated  cross sections are
given in the tables, expressed as ratios to the integrated cross sections 
$Z\sigma^{(\nu)}_{(\mathrm{Free})}$ (Eqn (\ref{W102a})) for $Z$ free electrons.

\section{Summary and conclusions}

The recently developed~\cite{Whitt2022} theory of scattering of low energy neutrinos and 
antineutrinos by atomic electrons, which uses the Bound Interaction Picture in
configuration space to fully implement the relationship between the neutrino helicities and 
the orbital and spin angular momenta of the atomic electrons, has been applied here to the 
scattering by the ground state systems Ne($1s^{2}2s^{2}2p^{6}$) and 
Ar($1s^{2}2s^{2}2p^{6}3s^{2}3p^{6}$). 

The energy spectra $d \sigma/dE_{f}$ of the ionization electrons produced in the scattering
of electron neutrinos and antineutrinos with energies 5, 10, 20 and 30 keV have been 
calculated. Results are also obtained for the integrated cross sections.
Screened point-Coulomb radial eigenfunctions have been used,
with the continuum state eigenfunctions calculated by direct integration of the Dirac equations.
The results here for Ne replace those given in~\cite{Whitt2022} which were calculated using 
some screening constants which underestimated the effects of screening in the 
inner subshells. The new spectra ratios for neon are larger than the original ratios by 
$\approx 0.03$ to $\approx 0.14$, with the greatest increases occurring at $E_{\nu_{i}}$ = 10 keV.

The calculated energy spectra show that binding effects increase with atomic number, 
decrease with increasing $E_{\nu_{i}}$ and, for each $E_{\nu_{i}}$, are greatest at the high electron 
energy end of the spectrum. The neutrino and antineutrino energy spectra are very similar, 
with small difference of $\alt 1\% $.

Binding effects are still very significant at $E_{\nu_{i}} = 30$ keV, the integrated spectra 
ratios being $\alt 0.6$ and $\alt 0.5$ for Ne and Ar respectively. Extension of the calculations
to higher neutrino energies to study the further decrease of binding effects would require 
a substantial increase in the maximum value of $|\kappa_{f}|$ used as the $\kappa_{f}$
convergence is very slow at higher energies, which is not practical. 

The largest uncertainty in the present calculations arises from the choice of the potential
$V(r)$ for the atomic electron, and the consequent form of the radial functions $g(r)$ and 
$f(r)$  used in the evaluation of the radial matrix elements~(\ref{W53}). The choice of a 
Coulombic potential with an effective nuclear charge $Z_{\mathrm{eff}}=Z-s_{i}$ to account
for electron screening effects is convenient
and should be accurate for the small values of $n$ and $n-l$ occurring in the ground states
of neon and argon~\cite{Bethe1957}. However, if increased accuracy is required, or extension to,
for example, xenon is considered, then self-consistent relativistic radial 
eigenfunctions~\cite{Kim1967,Smith1967} would be needed.

Relativistic self-consistent field theory includes~\cite{Kim1967} the effects of magnetic interactions between
electrons and retardation of the inter-electronic Coulomb repulsion, as described by the Breit operator.
This operator is treated as a first order perturbation to an unperturbed Hamiltonian which is a 
sum of the Dirac central field Coulombic Hamiltonians for each electron and the total Coulombic
inter-electronic interaction. 
The magnetic and retardation terms are of order $(v/c)^{2}$ compared to the inter-electronic repulsion
term, which is of order $(\alpha^{2}Z)m_{e}c^{2}$. Their contribution is therefore of order 
$(\alpha^{4}Z^{3})m_{e}c^{2}$ and increases strongly with $Z$. Detailed calculations are discussed
in section 7.2 of~\cite{Das1973}.

The scattering of dark matter (DM) particles by atomic electrons is a promising method for detecting
the particles forming the DM component of the Milky Way, and the theoretical treatment of this process
shares many features with that for neutrino scattering by atomic electrons. 
The response of argon and xenon targets has been studied by~\cite{Catena2020} by treating the atomic
electrons and DM particle as non-relativistic, with a general interaction constructed to satisfy
Galilean and rotational invariance. 
The DarkSide-50 experiment~\cite{Agnes2018} uses an argon target
and they have presented a non-relativistic analysis for a vector mediator based upon atomic and DM form factors.
The conversion of a sterile neutrino DM particle into an active neutrino via inelastic scattering
with an atomic electron has been considered by~\cite{GPS2022} using a non-relativistic 
second quantization formalism based upon a four-fermion coupling and a mediator which could be scalar, 
pseudo-scalar, vector or axial vector. 

The formalism developed in~\cite{Whitt2022}, which considers only
vector and axial vector couplings, could be readily adapted to the $Z$-mediated scattering of a fermionic
WIMP DM particle such as the neutralino $\chi$. The neutrino scattering tensor $L^{(\nu)}_{\beta \alpha}$
would be replaced by the appropriate tensor $L^{(\chi)}_{\beta \alpha}$ for the neutralino, 
the admixture of vector and axial vector contributions altered to accommodate the neutralino couplings, and changes
made to the normalization factors for the plane wave spinors. Generalization to scalar and pseudo-scalar
couplings is possible but would require substantial modifications to the formalism, especially to the
angular momentum matrix elements.

\begin{acknowledgments}
The author would like to thank Michael Meehan for assistance with the figures.
\end{acknowledgments}


\begin{thebibliography}{1}
\bibitem{Giunti2015}
C.~Giunti and A.~Studenikin,
Neutrino electromagnetic interactions: A window to new physics,
Rev. Modern Phys. \textbf{87}, 531 (2015) 

\bibitem{Jeong2021}
J.~Jeong, J.~E.~Kim and S.~Youn,
Electromagnetic properties of neutrinos from scattering on bound electrons in atoms,
Int. J. Mod. Phys A \textbf{36}, 2150182 (2021)

\bibitem{Kouzakov2014} K.~A.~Kouzakov and A.~I.~Studenikin,
Theory of neutrino-atom collisions: The history, present status, and BSM physics,
Advances in High Energy Physics \textbf{2014}, 569409 (2014)


\bibitem{Gounaris2002} G.~J.~Gounaris, E.~A.~Paschos and P.~I.~Porfyriadis, 
The ionization of H, He or Ne atoms using neutrinos or antineutrinos at keV energies,
Phys. Lett. B  \textbf{525}, 63 (2002)

\bibitem{Gounaris2004} G.~J.~Gounaris, E.~A.~Paschos and P.~I.~Porfyriadis,
Electron spectra in the ionization of atoms by neutrinos,
Phys. Rev. D \textbf{70}, 113008 (2004)


\bibitem{Chen2014}
J.-W.~Chen, H.-C.~Chi, K.-N.~Huang, C.-P.~Liu, H.-T.~Shiao, L.~Singh, H.~T.~Wong, C.-L.~Wu, and C.-P.~Wu,
Atomic ionization of germanium by neutrinos from \textit{ab initio} approach, Phys. Lett. B 
\textbf{731}, 159  (2014)

\bibitem{Chen2015}
Jiunn-Wei~Chen, Hsin-Chang~Chi, Keh-Ning~Huang, Hau~Bin~Li, C.-P.~Liu, Lakhwinder~Singh, Henry~T.~Wong, Chih-Liang~Wu, and Chih-Pan~Wu,
Constraining neutrino electromagnetic properties by germanium detectors,
Phys. Rev. D \textbf{91}, 013005  (2015)

\bibitem{Chen2017}
Jiunn-Wei~Chen, Hsin-Chang~Chi, C.-P.~Liu, and Chih-Pan~Wu,
Low energy electronic recoil in xenon detectors by solar neutrinos, Phys. Lett. B 
\textbf{774}, 656  (2017)


\bibitem{Whitt2022} I.~B.~Whittingham, Scattering of low enery neutrinos and antineutrinos by atomic electrons,
Phys. Rev. D \textbf{105}, 013008  (2022)

\bibitem{Furry1951} W.~H.~Furry, On bound states and scattering in positron theory, Phys. Rev. \textbf{81}, 115 (1951)

\bibitem{Whittingham1971} I.~B.~Whittingham, Incoherent scattering of gamma rays in heavy atoms, 
J. Phys. A:Gen. Phys. \textbf{4}, 21 (1971)

\bibitem{GPS2022} Shao-Feng Ge, Pedro Pasquini, and Jie Sheng,
Solar active - sterile neutrino conversion with atomic effects at dark matter direct detection experiments,
JHEP05 (2022) 088

\bibitem{Rose1961} M.~E.~Rose \textit{Relativistic Electron Theory} (New York, Wiley, 1961)

\bibitem{Das1973} T.~P.~Das \textit{Relativistic Quantum Mechanics of Electrons} (New York, Harper \& Row, 1973)

\bibitem{Bethe1957} H.~A.~Bethe and E.~E.~Salpeter \textit{Quantum Mechanics of One- 
and Two- Electron Atoms} (Berlin, Springer-Verlag, 1957)

\bibitem{Thomas1997} D.~Thomas \texttt{https://www.uoguelph.ca/chemistry \\
/thomas/index}

\bibitem{Schofield1973} J.~H.~Schofield,
Theoretical photoionization cross sections from 1 to 1500 keV,
Lawrence Livermore Laboratory report UCRL-51326 (1973),
\texttt{https://doi.org/10.2172/4545040}.

\bibitem{Kim1967} Yong-Ki~Kim,
Self-Consistent-Field Theory for Closed-Shell Atoms,
Phys. Rev. \textbf{154}, 17 (1967)


\bibitem{Smith1967} F.~C.~Smith and W.~R.~Johnson,
Relativistic self-consistent fields with exchange,
Phys. Rev. \textbf{160}, 136 (1967)

\bibitem{Catena2020} Riccardo Catena, Timon~Emken, Nicola~A.~Spaldin, and Walter~Tarinto,
Atomic responses to general dark matter-electron interactions,
Phys. Rev. Res. \textbf{2}, 033195 (2020)

\bibitem{Agnes2018} P.~Agnes \textit{et al.} (DarkSide),
Constraints on Sub-GeV Dark-Matter Electron Scattering from the DarkSide-50 Experiment,
Phys. Rev. Lett. \textbf{121}, 111303 (2018)



\end{thebibliography}
\end{document}